\documentclass[fleqn,usenatbib]{Definitions/mdpi} 
\firstpage{1} 
\pubvolume{1}
\issuenum{1}
\articlenumber{0}
\pubyear{2023}
\copyrightyear{2023}
\datereceived{ } 
\daterevised{ } 
\dateaccepted{ } 
\datepublished{ } 
\hreflink{https://doi.org/} 

\Title{Impact of AGN feedback on the dynamics of gas; A Review across Diverse Environments}

\TitleCitation{Title}

\Author{ Mojtaba Raouf  $^{1,2,*}$\orcidA{} ,
 Mohammad Hossein Purabbas $^{2,3}$,
 Fatemeh Fazel Hesar $^{1,4}$
}

\AuthorNames{Mojtaba Raouf, Mohammad Hossein Purabbas and Fatemeh Fazel Hesar}

\AuthorCitation{Raouf, M.; Purabbas , M. H.; Fazel Hesar, F.}

\address{%
$^{1}$ \quad Leiden Observatory, Leiden University, P.O. Box 9513, 2300 RA Leiden, Netherlands \\
$^{2}$ \quad School of Astronomy, Institute for Research in Fundamental Sciences (IPM), Tehran, 19395-5746, Iran \\
$^{3}$ \quad  Department of Physics, Faculty of Sciences, University of Birjand, Shahid Aviny Street, Birjand, Iran \\
$^{4}$ \quad Eurospacehub academy \& LUNEX, ESA BIC, Noordwijk, Netherlands}
\corres{Correspondence: mojtaba.raouf@gmail.com, raouf@strw.leidenuniv.nl}

\abstract{
This review examines the relationship between black hole activity and kinematic gas-star misalignment in brightest group galaxies (BGGs) with different merger rates. 
The formation history of galaxy groups is assessed through "age-dating" as an indicator of distinct major mergers involving the BGG. BGGs within groups characterized by a higher frequency of major mergers are more likely to host active SMBHs.
A consistent correlation is identified between the level of black hole activity, as indicated by the 1.4 GHz and 325 MHz radio emissions, and the degree of kinematic misalignment between the gas and stellar components in BGGs.
In dynamically fossil groups, where black hole accretion rate is relatively ($\sim 1$ dex) lower due to the lack of recent ( $\leq 1$ Gyr) major mergers, there is reduced ($\sim$ 30\%) misalignment between the gas and stellar components of BGGs compared to non-fossil groups. 
Additionally, the study reveals that BGGs in non-fossil groups show higher levels of star formation rate and increased occurrence of mergers, contributing to observed color differences. Exploring the properties and dynamics of the gas disc influenced by mechanical AGN feedback through hydrodynamic simulations suggests that AGN wind-induced effects further lead to persistent gas misalignment of the disk around the supermassive black hole.
}


\begin{document}

\section{Introduction}

The presence of super massive black holes (SMBH) at the centers of massive galaxies provides a mechanism for the existence of Active Galactic Nuclei (AGN), which can suppress excessive star formation and regulate galaxy growth \cite{silk1998,binney1995}. AGN feedback, through processes such as heating the surrounding gas with AGN jets, plays a crucial role in quenching massive galaxies and controlling the growth of SMBHs \cite{Kondapally2023}. This feedback has a direct influence on the gas content of the galaxy, resulting in negative/positive feedback on star formation \cite{silk2005ultraluminous,Silk2024,gaibler2012jet,santini2012enhanced,bieri2015playing}. The role of AGN feedback in shaping galaxy properties is supported by numerous studies \citep{benson2003galaxy,di2005energy,bower2006breaking,croton2006many,sijacki2007unified,cattaneo2009role,fabian2012observational}.

Brightest groups galaxies (BGGs), at the core of galaxy groups, serve as windows into the intricate processes governing galaxy assembly, star formation, and the interplay between SMBHs and their host galaxies \citep{einasto2024galaxy,Raouf2021}. In a group, galaxy interactions shape galaxy evolution by leaving observable marks like tidal features and kinematic perturbations in stars and gas \citep{haynes1984influence,bertola1992gaseous,sancisi2008cold}. A remarkable perturbation is the stellar-gas misalignment, where stars and gas rotate in different (in case of conter-rotation, opposite) directions due to their various (opposite) angular momentum \citep{kannappan2001broad} . 
A significant deviation exceeding 30$^{\circ}$ in the kinematic position angle (PA) between gas and stars can arise from various sources, both internal and external to the galaxy. These sources may include mergers between galaxies, gas accretion from neighboring galaxies, or the influence of black hole activity, such as AGN feedback 
\citep{bertola1984stellar,davis2016depletion,raimundo2023}.

Dynamically fossil groups, which lack recent group-scale mergers\footnote{While group-group mergers can potentially lead to increased merging in the overall galaxy population over time, they do not necessarily guarantee that mergers involving the dominant galaxy will occur.} and major galaxy mergers, are particularly suitable for such studies \citep{ponman1994possible, jones2003}. The dynamical state/age of galaxy groups can be characterized by indicators such as the luminosity gap and the offset between the BGG and the luminosity centroid, with the luminosity gap being a key factor \citep{Raouf2014}. A large luminosity gap suggests the absence of recent major mergers that could trigger cold mode accretion. 
In dynamically fossil groups, where the intergalactic medium (IGM) reaches its peak density at the bottom of the group/cluster potential well, an AGN is subject to hot gas accretion. 
In fact, the observational data is much clearer for non-fossil systems (with the low luminosity gap), showing that many of the most powerful radio galaxies in groups in the local universe are the BGGs of dynamically relaxed systems with cooling cores, fuelled by gas cooling from the ICM, not from galaxy mergers. 
On one hand, \citep{Khosroshahi2017} found a combination of the low luminosity gap (an indicator for internal event and cold accretion) and the large BGG offset from the centre (an indicator for external event and hot accretion) in a group, is the key driver behind the observed high AGN activity probed by the radio emission. 
The presence of nuclear gas that can be accreted by a central supermassive black hole is considered crucial for fueling and powering AGN  \citep{shlosman1990fuelling}. 
On the other hand, the prevalence of misalignments in early-type galaxies (which is mostly located at the center of groups and clusters), along with the possibility of aligned stellar-gas kinematics resulting from external gas accretion depending on the interaction geometry and galaxy morphology, indicates a substantial contribution of external accretion to the total gas content in these galaxies \cite{Ristea2022} , which in turn affects the availability of gas for black hole fuelling. Based on the research conducted by \cite{Bryant2019}, they considered the possibility that kinematic misalignments in clusters could be attributed to gas stripped from the galactic disk as the galaxy moves through the intra-cluster medium. This process would lead to a gradient in the gas velocity field in the direction of infall. However, it is important to note that after a thorough examination of the same dataset, \cite{Ristea2022} found that only one galaxy, representing less than 1\% of the sample, supported a misalignment origin due to ram-pressure stripping. These studies highlight the complexities involved in understanding the causes of kinematic misalignments in the cluster environment. While gas stripping through hydrodynamic effects can contribute to misalignments in some cases, the prevalence of such misalignments in clusters is not well-supported by the available evidence.

The alignment or misalignment of gas and stars in galaxies, along with the influence of external gas accretion, plays a crucial role in understanding the fuelling of SMBHs and the AGN.
Theoretical studies have proposed that the presence of counter-rotating or significantly misaligned structures facilitates gas inflow and potentially contributes to the fuelling of supermassive black holes in galaxies of various types \citep{thakar&ryden1998,vandevoort2015,negri2014,capelo2017,taylor2018}   .
Recently, a hydrodynamic simulation of idealised gas disc around SMBH suggested that the stability of the disk's PA across different radii can be attributed to the mechanical feedback from the AGN, indicating a direct influence of AGN activity on the kinematic properties of the molecular gas \cite{Raouf2023}. 

This comprehensive review primarily focuses on results derived from simulations, offering valuable insights into the impacts of black hole activity on the kinematics of gas during the evolution of galaxies. The paper is structured into key sections that cover essential aspects, starting with an examination of the dynamical state of galaxy groups, followed by an exploration of the complex interplay between black hole evolution and their actions, and finally delving into the dramatic narratives of major merger history.
In addition to the simulation-based findings, this review incorporates observational studies that further enhance our understanding of the subject matter. These studies shed light on various important aspects, including the variations in radio emissions from BGGs, the influence of galaxy groups' dynamical states on their stellar populations, and the intriguing phenomenon of kinematic gas-star misalignment.
Furthermore, by employing hydrodynamic simulations, we present the fascinating phenomena of gas misalignment in AGN-dominated galaxies, specifically focusing on an idealized gas disc around the supermassive black hole.
Each segment of this review contributes a valuable piece to the puzzle, enriching our understanding of the complex interplay between black hole activity, kinematic misalignment, and the environments of galaxies. By combining simulation results and observational insights, we aim to provide a comprehensive and nuanced perspective on this captivating field of research.

\begin{figure}
    \centering
    \includegraphics[width=0.49\linewidth]{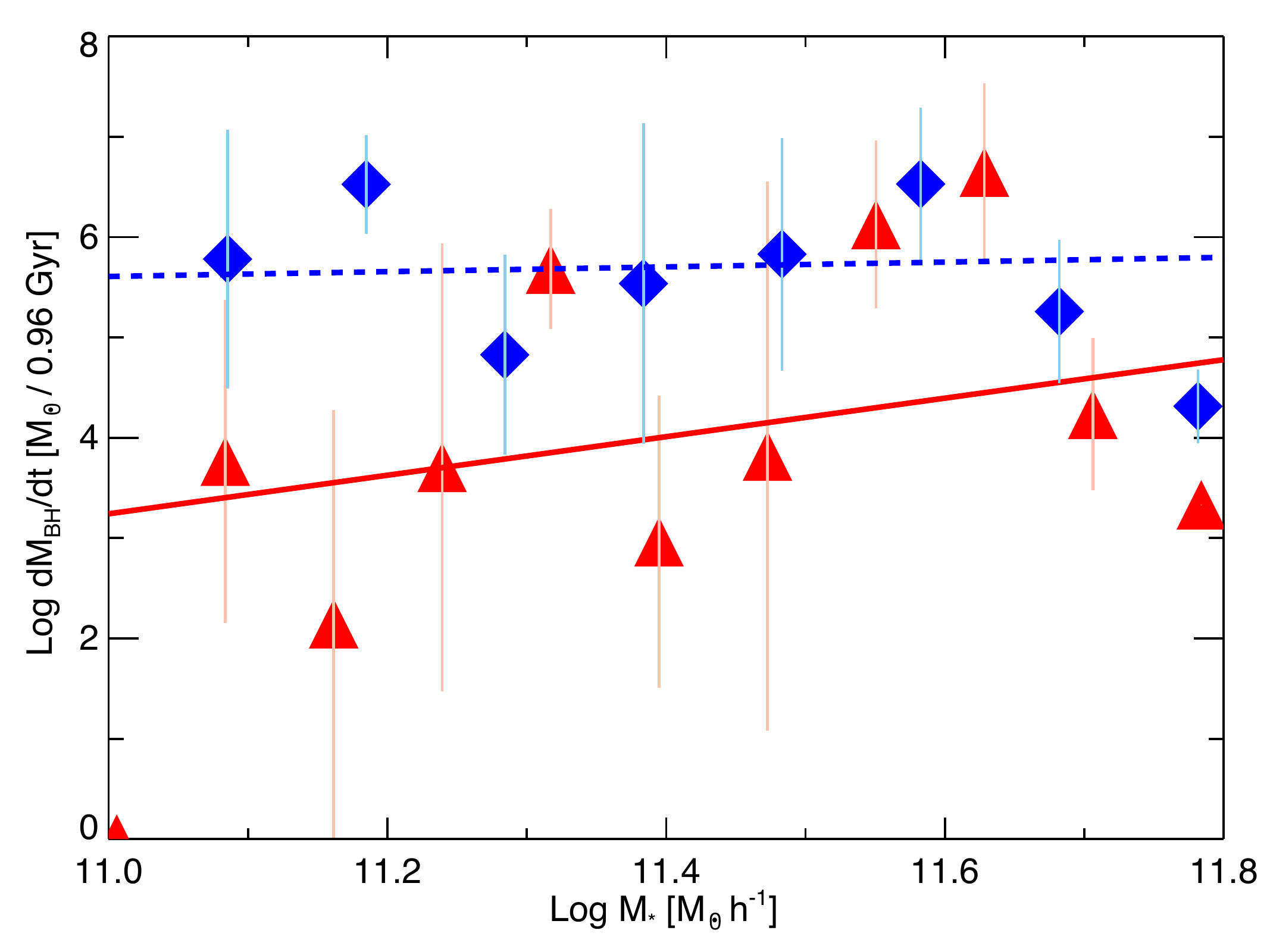}
    \includegraphics[width=0.49\linewidth]{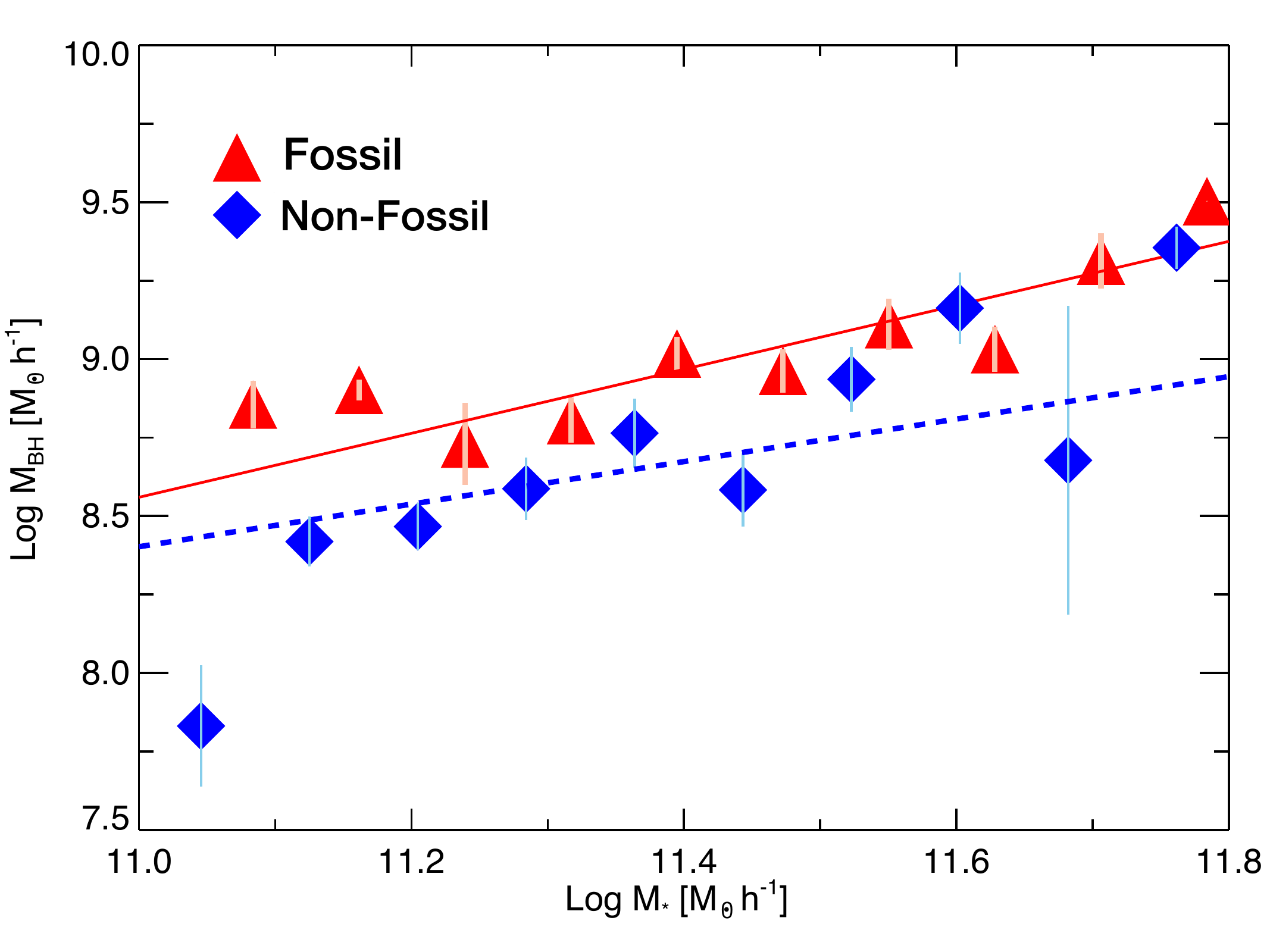}
    \caption{\textbf{Left}: The black hole accretion rate ($\dot M_{BH}$) as function of stellar mass for BGGs, comparing dynamically fossil groups (red) against BGGs in dynamically non-fossil groups (blue). The red line and blue dashed line show the linear regressions to the fossil and non-fossil systems, respectively. \textbf{Right}: The black hole mass ($M_{BH}$) as function of stellar mass for BGGs in fossil and non-fossil groups. the BGG in fossil group generally possessing larger black holes than those in non-fossil groups, reflecting the influence of the group's dynamical history on black hole growth. The error bars in all panels are based on the standard deviation
over the mean.}
    \label{fig:BH accretion rate&mass}
\end{figure}

\section{Perspectives from cosmological simulations}
\subsection{Age Dating Galaxy Groups}

 Fossil galaxy groups refer to systems that have undergone relatively few recent mergers or interactions \cite{Khosroshahi2006}. These groups are characterized by a more stable and virialized state, where the member galaxies and dark matter halos have had sufficient time to settle into a dynamically old state/age configuration. 
 In fossil groups, galaxies are more centrally concentrated, which suggests that a greater number of member galaxies are located near the center rather than evenly distributed throughout the group \cite{Raouf2014}. 
 Additionally, the X-ray emitting hot gas in these groups exhibits a smoother and more symmetric distribution \citep{Khosroshahi2004}.
In contrast, non-fossil galaxy groups are systems that have experienced recent merger events or interactions with other groups or galaxies. Due to ongoing interactions, the member galaxies in non-fossil groups may exhibit more chaotic motions, and the spatial distribution of galaxies can be more dispersed \citep{Aguerri2021}. The X-ray emission from the hot gas in non-fossil groups may show signs of substructures or asymmetries \cite{Khosroshahi2004}. Such differences in the dynamical state of galaxy groups significantly impact the stellar populations and gas kinematics of the BGGs, with non-fossil groups showing significantly bluer NUV-r colors and higher star formation rates compared to their fossil counterparts \citep{Raouf2019, Raouf2021}. 
The advantage of \cite{Raouf2014} study is its use of optical data to distinguish  between these fossil and non-fossil groups. By analyzing various measurable parameters derived from optical observations of the galaxy groups, the study aims to distinguish between groups with different merger histories \cite{Raouf2018}. 
Several studies explore the accuracy of photometric, spectroscopic, and X-ray proxies in assessing the dynamical state of galaxy groups \cite{Parekh2015,Khosroshahi2004}. In the current era of extensive surveys, having a reliable and cost-effective method to characterize the dynamical state of halos would greatly benefit statistical studies focused on understanding the influence of the environment on galaxy properties. By utilizing the luminosity gap and BGG offset approach, \citep{Haghighi2020} show that such an alternative approach that mitigates the need to rely exclusively on X-ray data or other observational techniques, which may be limited by sample size or coverage. 

The review study of fossil groups of galaxies offers a distinct perspective on the processes influencing galaxy evolution, providing a transitional picture \cite{Aguerri2021}. Fossil groups are characterized by a large luminosity gap ($\Delta M_{12} > 2$ mag) between the BGG and the other group members along with relatively high X-ray luminosity ($L_{X,bol}\approx 10^{42} $~h$_{50}^{-2}$ ~erg~s$^{-1}$) \citep{jones2003, ponman1994possible}, indicating a long period of dynamical relaxation \citep{Khosroshahi2016,Farhang2017}. Moreover, by adding the offset from the group luminosity centroid, we can considerably improve the accuracy and precision of \textit{age-dating} techniques for optically (without X-ray criterion) selection fossil groups of galaxies \citep{Raouf2014}. 
In this review, we refer to the term "fossil" to describe an "evolved" group that is more commonly observed and represents an "older" system in simulations, indicating its forming earlier. The non-fossil group, in contrast, represents the "evolving" groups, which are younger and formed at a later stage.

\subsection{black hole activity and formation history}

 The BGGs in fossil groups typically harbor more massive black holes, underscoring a correlation between the historical mass assembly of group halos and the growth of SMBHs. Figure \ref{fig:BH accretion rate&mass} provides a visual representation of the variance in black hole accretion rates and mass within BGGs hosted by fossil and non-fossil groups using cosmological simulations \citep{Raouf2016}. The BGG in the dynamically fossil galaxy groups display a lower accretion rate (left panel) compared to the BGG in dynamically non-fossil groups. At the same time the black hole mass (right panel) in the BGGs dominating the dynamically fossil groups is larger than in the BGGs of dynamically non-fossil or young groups. This means that the mass assembly history of the host group (halos) has a significant impact on the activity and growth of BGG's SMBH. The BGGs of fossil systems, seem to be very efficient in black hole growth by consuming the gas which could have been generally found with a higher density in early stages of the halo formation \cite{Raouf2016}.

\begin{figure}
    \centering
    \includegraphics[width=0.33\linewidth]{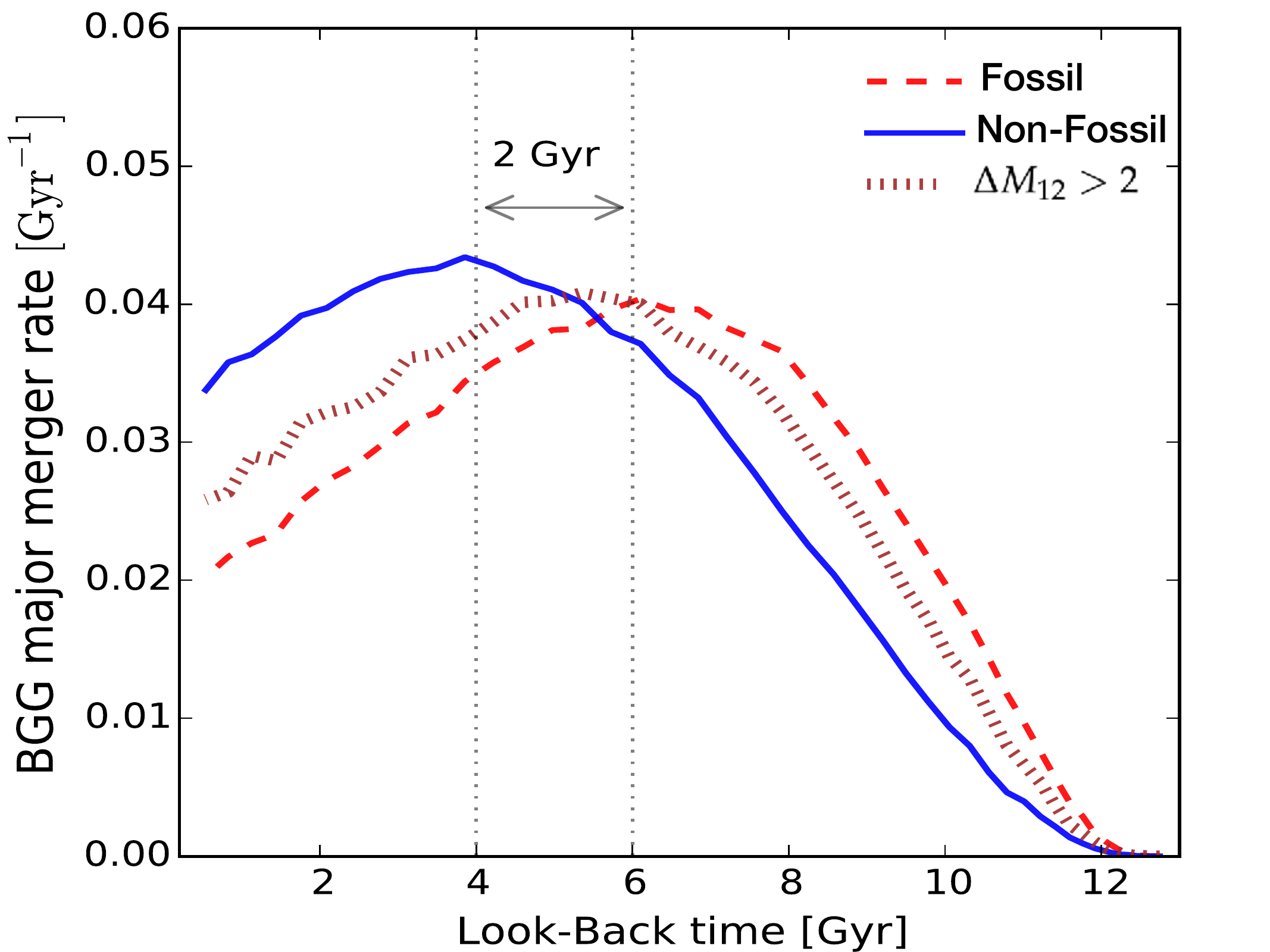}
    \includegraphics[width=0.33\linewidth]{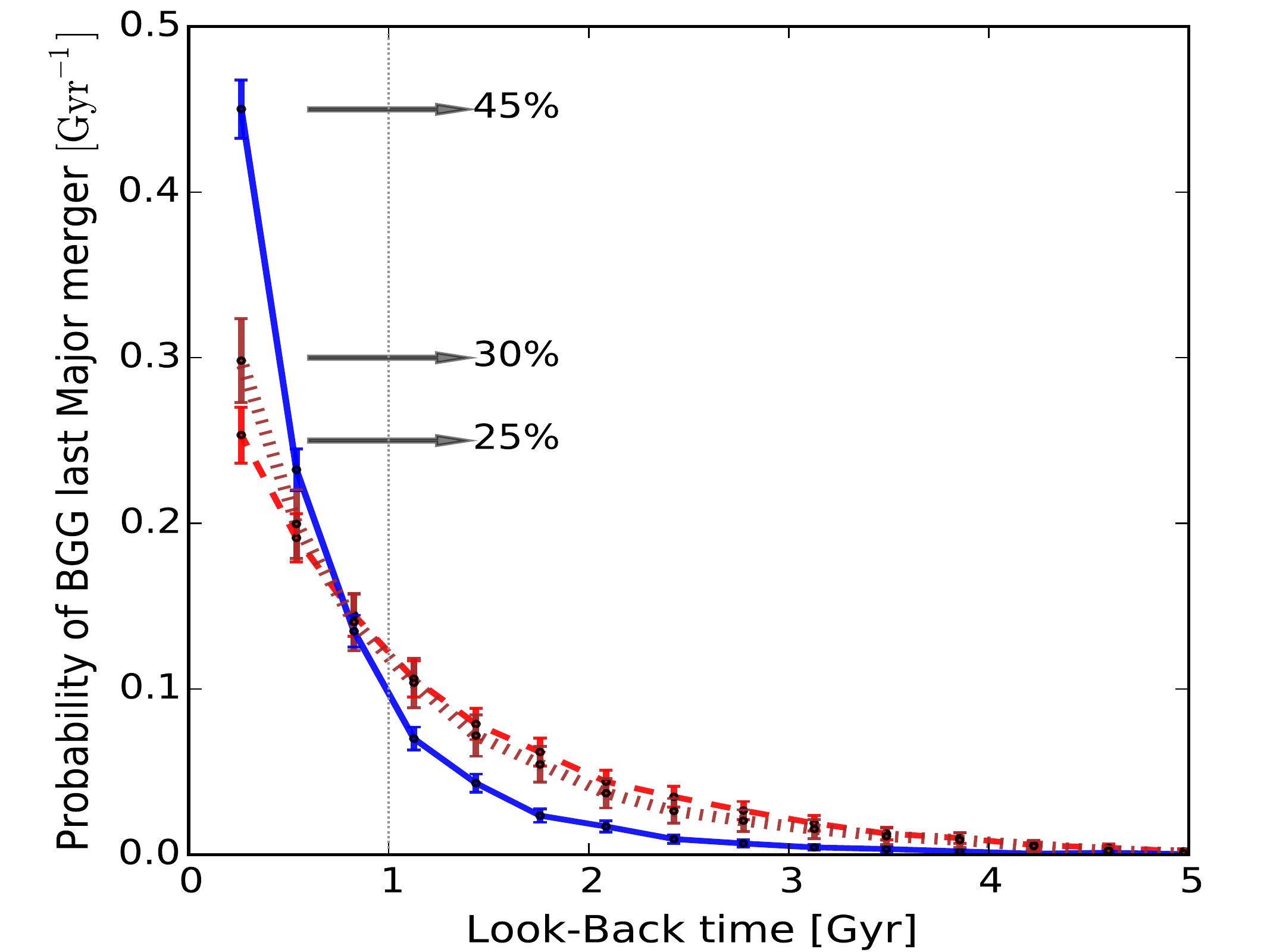}
    \includegraphics[width=0.32\linewidth]{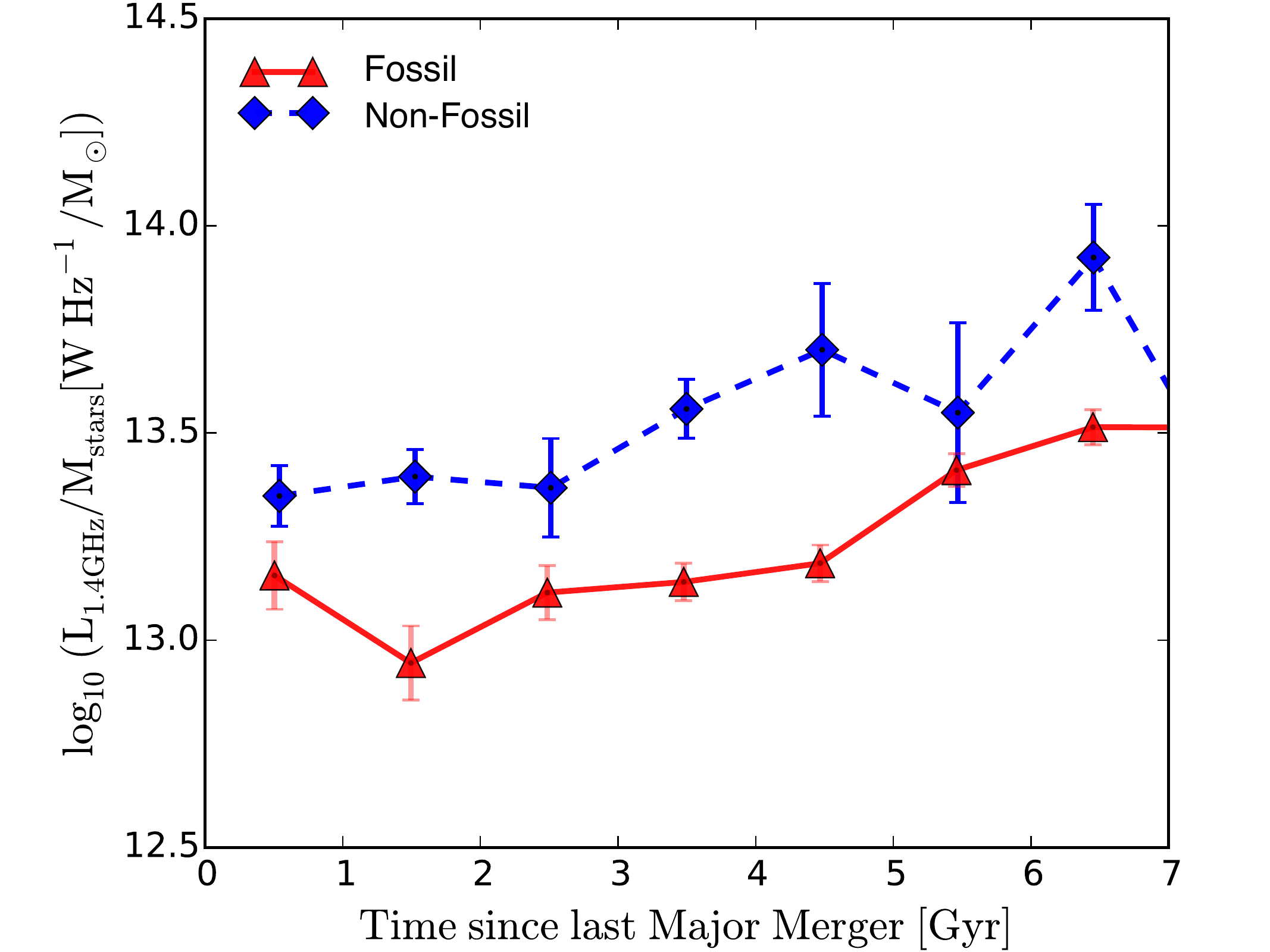}
    \caption{\textbf{Left}: The major merger rates of the BGG in three categories of groups using Radio-SAGE galaxy formation model: dynamically fossil (red dashed line) and non-fossil (blue line) groups of galaxies along the fossil ($\Delta M_{12}>2$) groups (brown dotted line), for halo masses above $10^{13} M_{\odot}$. Major mergers are characterized by mergers with \textit{$m_{1}/m_{2}<3$}. As seen, the BGGs of fossil systems show a higher probability for a major merger at earlier epochs compared with non-fossil systems (about 2 Gyr in peak).  \textbf{Middle}: Comparison of the distributions of time since the last major merger between the three group categories. The error bars represent Poisson uncertainties. The most recent major merger event for BGGs in non-fossil groups occurred more recently compared to the BGGs in high gap only and fossil groups, respectively.
. \textbf{Right}: Radio loudness (luminosity at 1.4 GHz divided by stellar mass) of BGGs at z=0 versus the time since the last major merger suffered by the BGG, in subsamples of groups defined by the epoch of their last major merger. The red triangles and blue diamonds show the medians and standard deviation over mean  error for BGGs of fossil and non-fossil groups, respectively}.
    \label{fig:2}
\end{figure}

Through the growth history of fossil and non-fossil halo, we found a significant time difference ($\sim$ 2 Gyr) in their peak merging activity \cite{Raouf2018}. As shown in left panel of Figure \ref{fig:2},  BGGs in non-fossil groups had their merger peak about 2 billion years (Gyr) later than BGGs in fossil groups. As expected, galaxies in the two types of groups have different merger histories. 
Specifically, our analysis reveals that non-fossil groups, on average, have experienced last major mergers with a frequency approximately $\sim$ 2 times higher within the past 1 Gyr compared to the BGGs in fossil groups (middle panel). It is important to note that these last major mergers refer to the overall history of the groups and may not necessarily correspond to immediate major mergers involving their galaxies within the specified 1 Gyr time frame.
This relationship is further supported by the increased radio brightness ($L_{1.4 GHz}/M_{star}$) illustrate in the right panel of Figure \ref{fig:2} utilising radio semi-analytic galaxy evolution model (Radio-SAGE; \cite{Raouf2017,Raouf2019a}). 
In contrast, BGGs in fossil groups exhibit an earlier peak in merging activity and demonstrate lower levels of radio brightness, even when considering the same duration since the last major merger \cite{Raouf2018}. 
 
\section{Perspectives from observational surveys}
\subsection{Radio emissions of BGGs}
\begin{figure}
    \centering
    \includegraphics[width=0.49\linewidth]{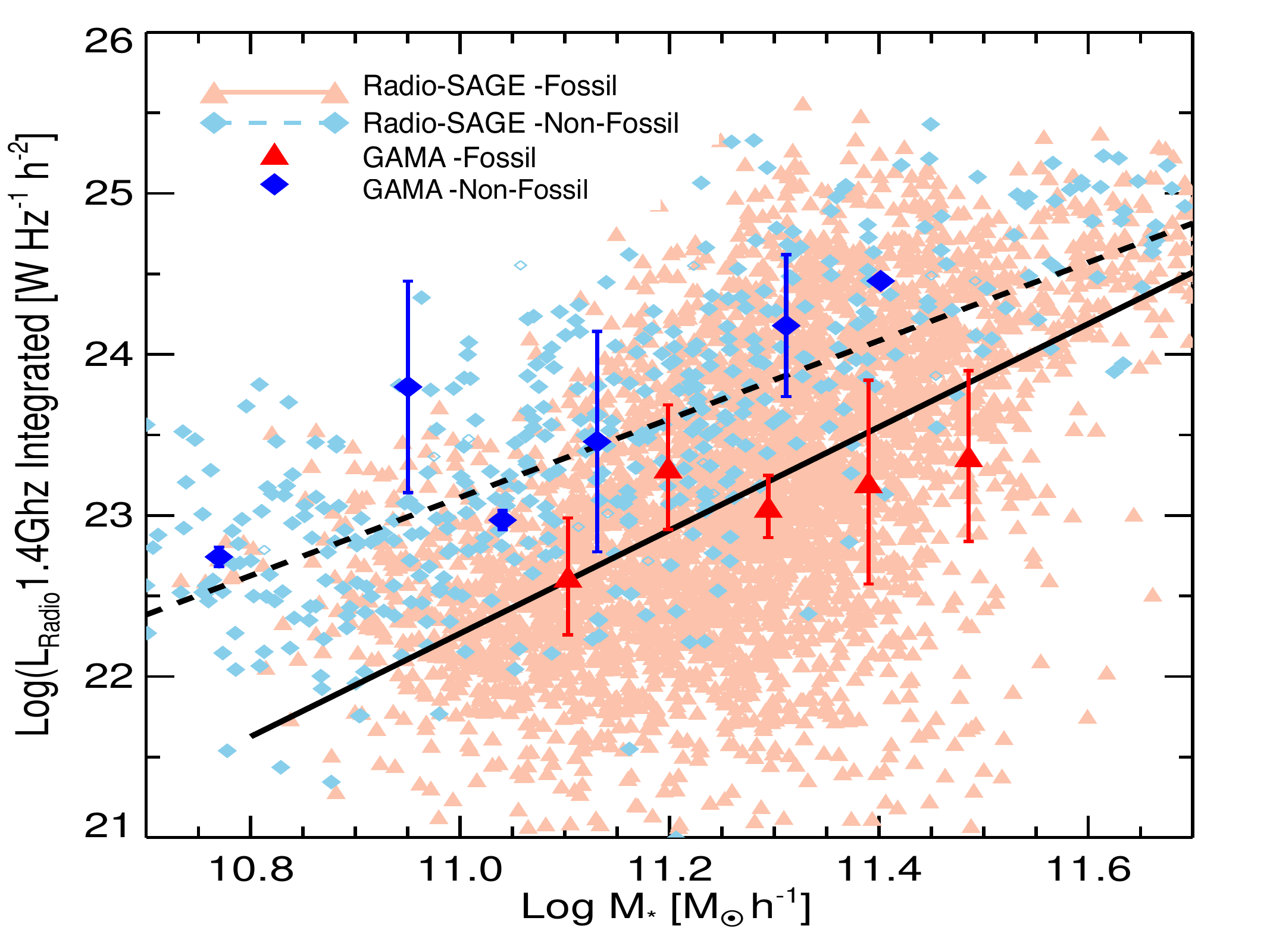}
    \includegraphics[width=0.49\linewidth]{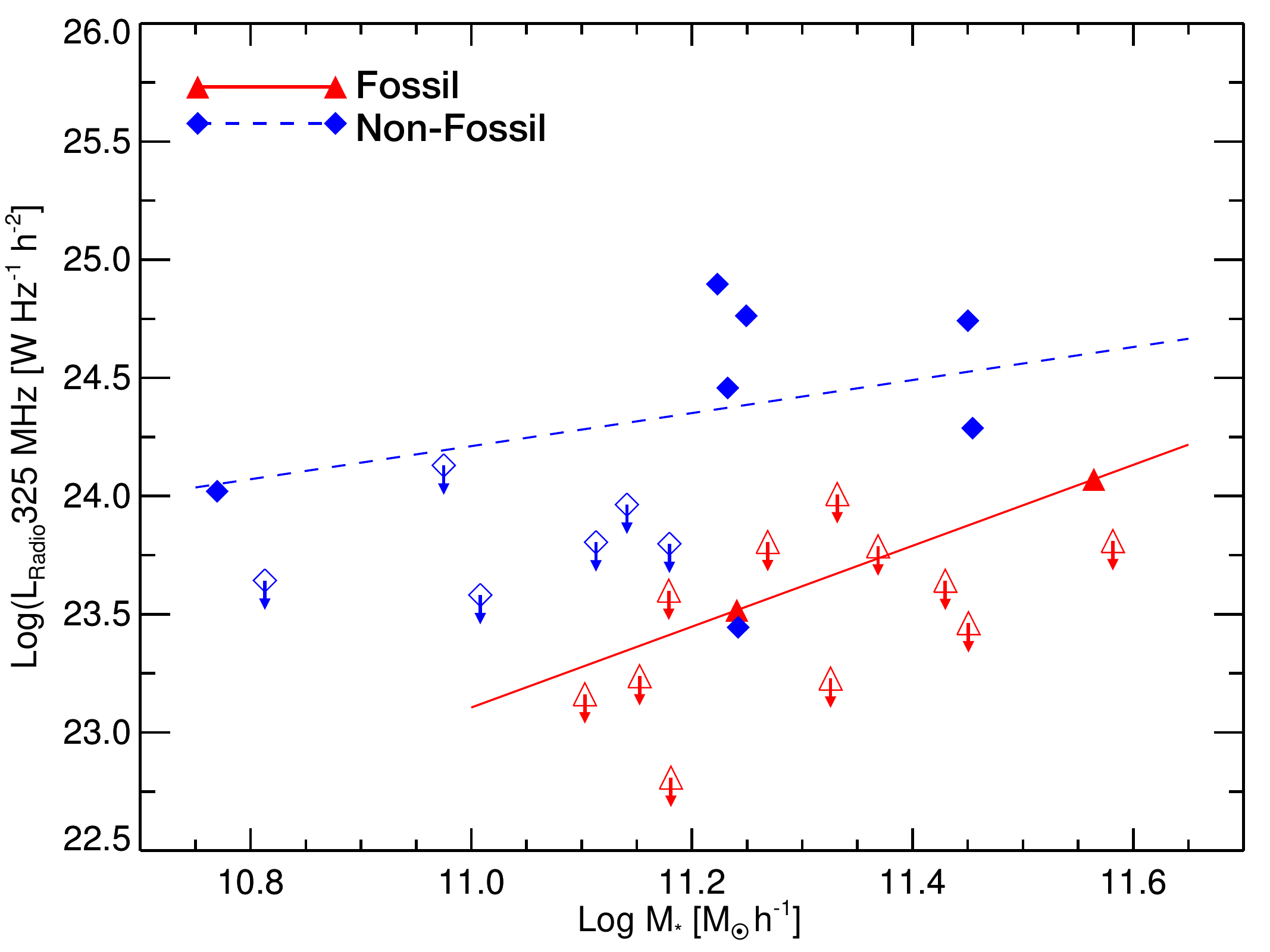}
    \caption{\textbf{Left}: The radio emission at 1.4 GHz for the BGG in fossil groups (red) and non-fossil groups (blue) as function of stellar mass. The radio luminosity refers to the integrated flux densities obtained from the VLA FIRST catalog. The bold symbols indicate the average value over the bin. We overlay the central galaxies in fossil (light red triangles) and unrelxed (sky blue diamonds) galaxy groups corresponding to different stellar masses as a function of the 1.4 GHz radio luminosity predicted by Radio-SAGE galaxy formations model. The solid and dashed-lines represent linear fits to the model data points for the BGGs in fossil and non-fossil galaxy groups, respectively. The majority of BGGs in dynamically non-fossil groups are radio-loud (i.e $L_{1.4 GHz} > 10^{23}\ W Hz^{-1} h^{-2}$).
    \textbf{Right}: 325 MHz radio power of the BGGs in fossil (filled red) and non-fossil (filled blue) groups using GMRT observations. An upper limit is given for the undetected fossil (open red) and non-fossil (open blue) BGGs.}
    \label{fig:3}
\end{figure}

Several observational studies consistently demonstrate a notable increase in AGN activity within merging galaxies \cite{Alonso2007,Weston2017}.
As shown in Figure \ref{fig:3} for 1.4 GHz (left panel) and 325 MHz (right panel) radio emission, the study by \cite{Khosroshahi2017} indicates that in the dynamically fossil environments, BGGs exhibit a diminished level of radio activity ($\sim$ 1 dex), indicating a relatively quiet phase of AGN activity. Our findings reveal that the proportion of radio-loud brightest group galaxies in observed dynamically non-fossil groups is approximately twice that found in dynamically fossil groups. 
Incorporating the Radio-SAGE\footnote{https://github.com/mojtabaraouf/sage} \citep{Raouf2017} model's findings into the discussion on radio emissions in BGGs, it becomes evident that the refined modeling of AGN feedback significantly contributes to our understanding of AGN activity and its various impacts. The model's advanced approach to simulating AGN jets and cooling processes under different dynamical states of galaxy groups complements the observed radio emission patterns in BGGs. This shows the complex connection between the dynamical state of the galaxy group, the behavior of the central black hole, and the resultant AGN feedback, elucidating the varied radio emission properties of BGGs in different group environments. The successful reproduction of this observation has been achieved through the utilization of a Radio-SAGE model. This model enables us to investigate the different factors contributing to these findings.
The consistent agreement between observations and Radio-SAGE model predictions indicates a higher rate of accretion\footnote{The stable gravitational potential within fossil groups reduces the accretion rate onto the central supermassive black hole compared to more dynamically non-fossil environments.} for the central black hole hosted by the BGG in dynamically non-fossil galaxy groups compared to those hosted by dynamically fossil galaxy groups at a given stellar mass.
\citep{Pasini2021} also emphasizing the influence of group dynamics on AGN feedback mechanisms of BGGs through the study of kinematic properties and scaling relations of radio galaxies in galaxy groups. \cite{Miraghaei2014} utilized a small sample of fossil groups to investigate the radio emission properties of the most luminous galaxies in these dynamically fossil halos. Their findings indicated that these galaxies exhibit a lower level of radio emission at 610 MHz and 1.4 GHz. 
The findings of these studies suggest that mergers, which are the primary events driving the formation of a large luminosity gap, are likely the main factor contributing to the differences observed in the radio properties.
 
\begin{quote}

\end{quote}

\subsection{Stellar population of BGGs}

\begin{figure}
    \centering
    \includegraphics[width=0.49\linewidth]{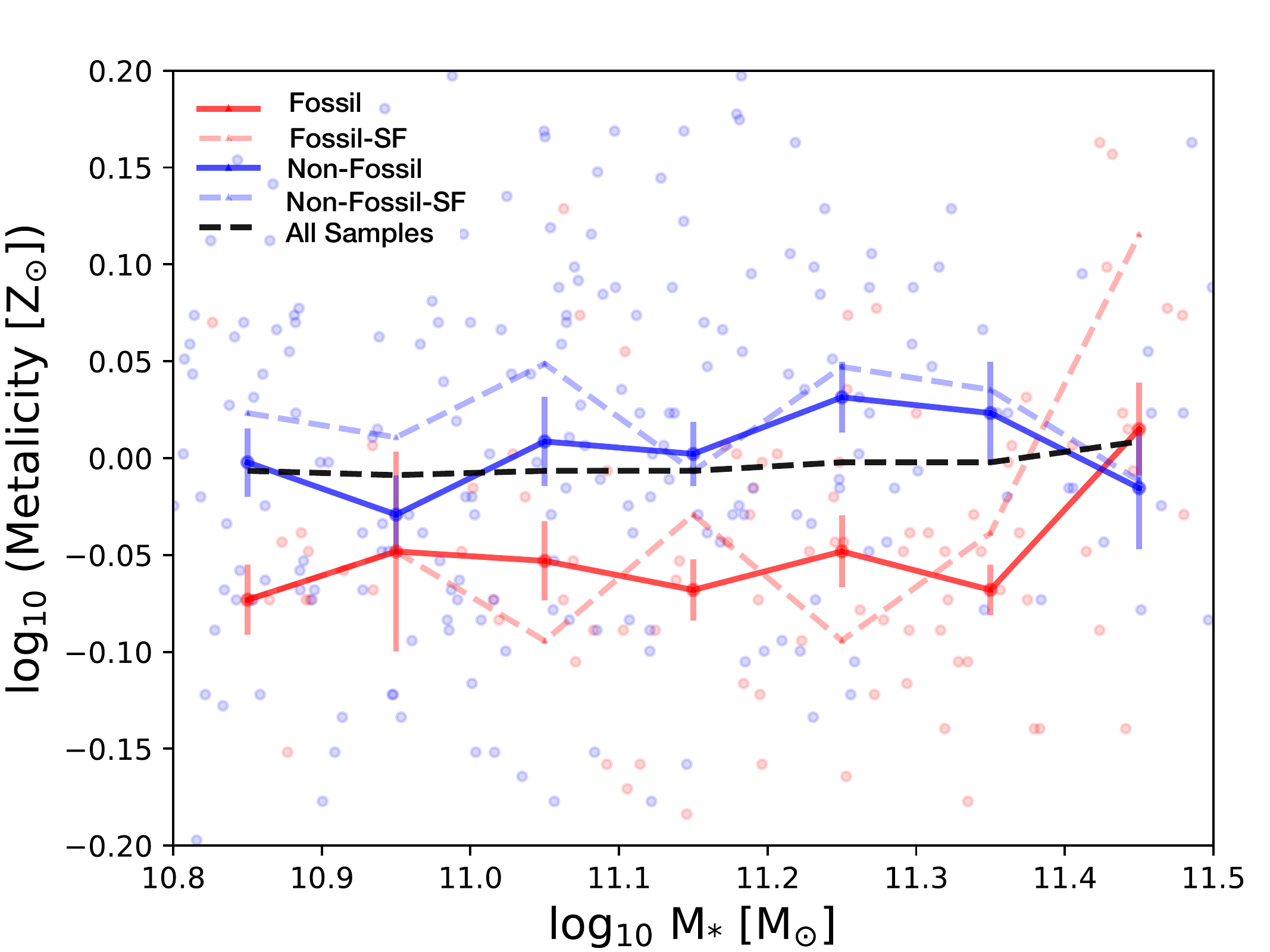}
    \includegraphics[width=0.49\linewidth]{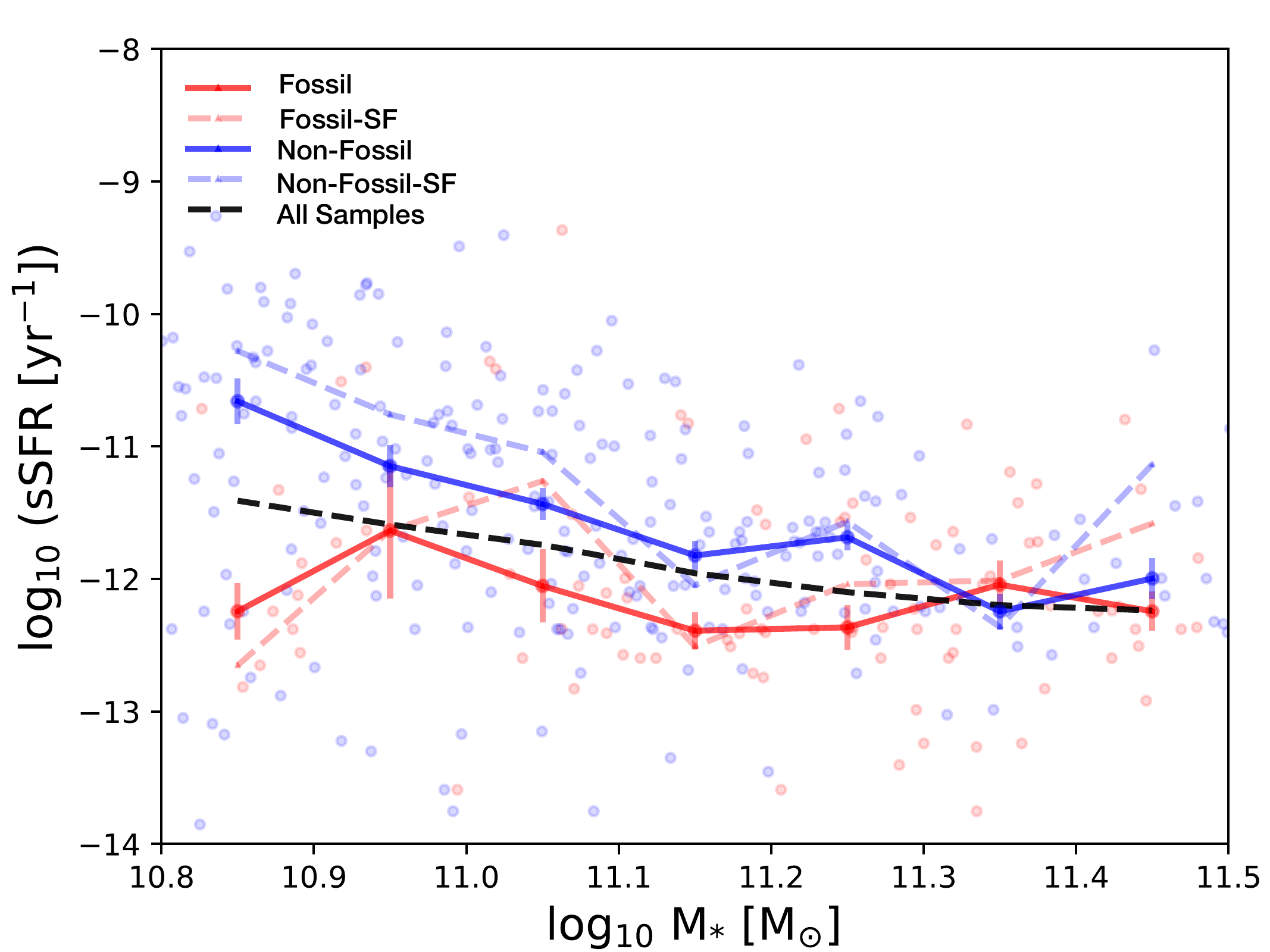}
    \caption{The variation in metallicity (left) and sSFR (right) of BGGs in relation to their stellar mass with the medians and  standard deviation over mean errors, for fossil and non-fossil groups with specified on all and star forming (SF, light-colors) BGGs selected by BPT diagram. The dashed black line in each panel is the metallicity and sSFR for the full sample. BGGs in fossil groups (red line) generally have lower metallicity than those in non-fossil groups (blue line) while the median sSFR is higher for BGGs in non-fossil groups compared to those in fossil groups.}
    \label{fig:4}
\end{figure}

Figure \ref{fig:4} shows the metallicity and specific star formation rate (sSFR) derived from SED fitting as a function of stellar mass for a samples of Galaxy And Mass Assembly (GAMA) observations \cite{Raouf2019}. 
As shown in the right panel, BGGs in non-fossil groups tend to show higher levels of star formation rate (SFR) compared to those in fossil groups. 
Also, the BGGs in fossil groups tend to have a higher percentage of red NUV-r colours (with higher precentage of NUV-r > 4.5) considering a fixed Sérsic index and dust mass, again indicating lower star formation rate.
This suggests that the increased occurrence of mergers in non-fossil groups contributes to the observed NUV-r color differences. 
In the left panel BGGs in non-fossil groups tend to have higher stellar metallicities ($\sim$0.05 dex) compared to those in fossil groups. This could be attributed to the mass and metal enrichment history of the building blocks from which the groups formed.

\subsection{ The Kinematics of BGG}

\begin{figure}
    \centering
    \includegraphics[width=0.49\linewidth]{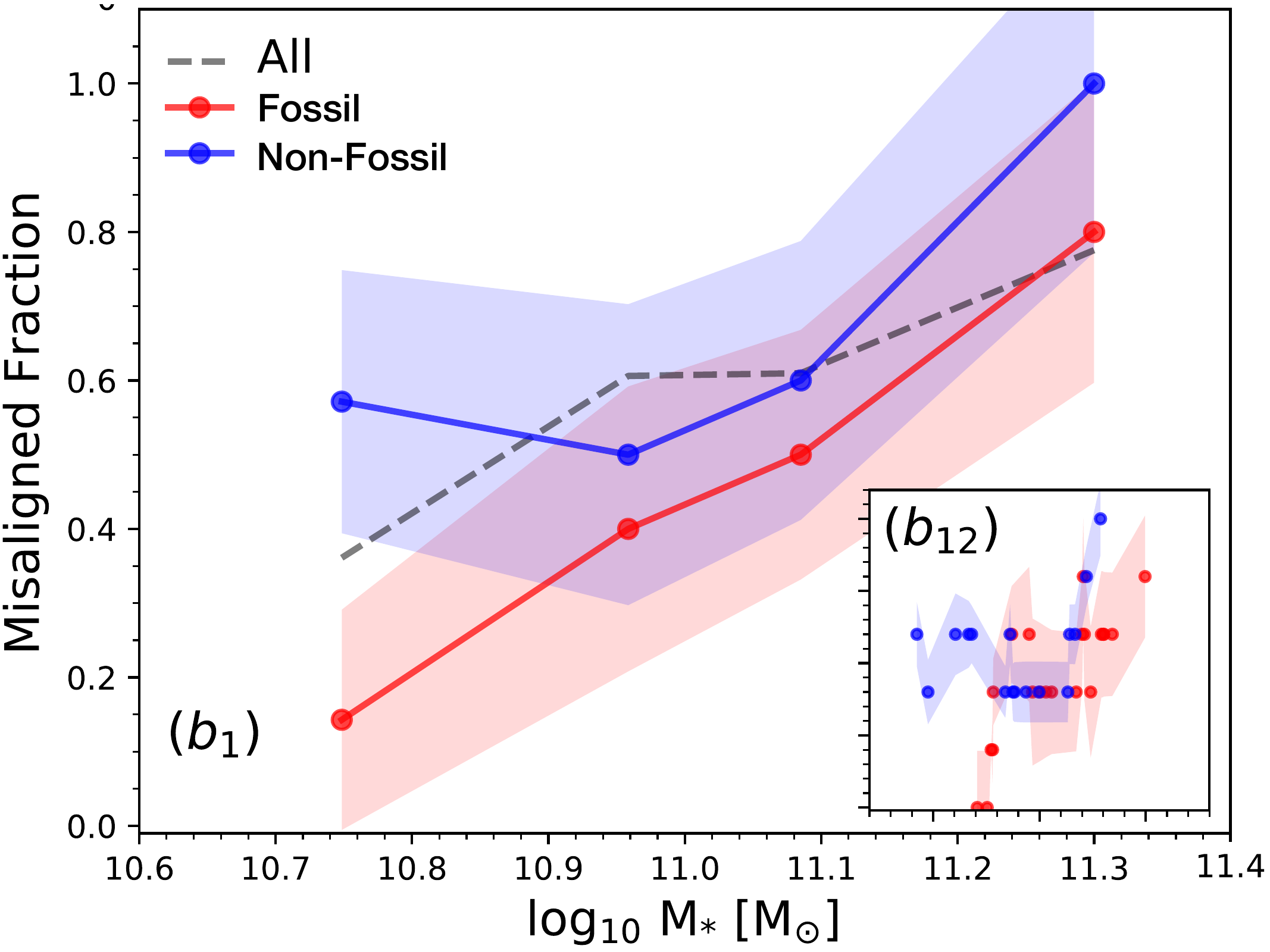}
    \includegraphics[width=0.49\linewidth]{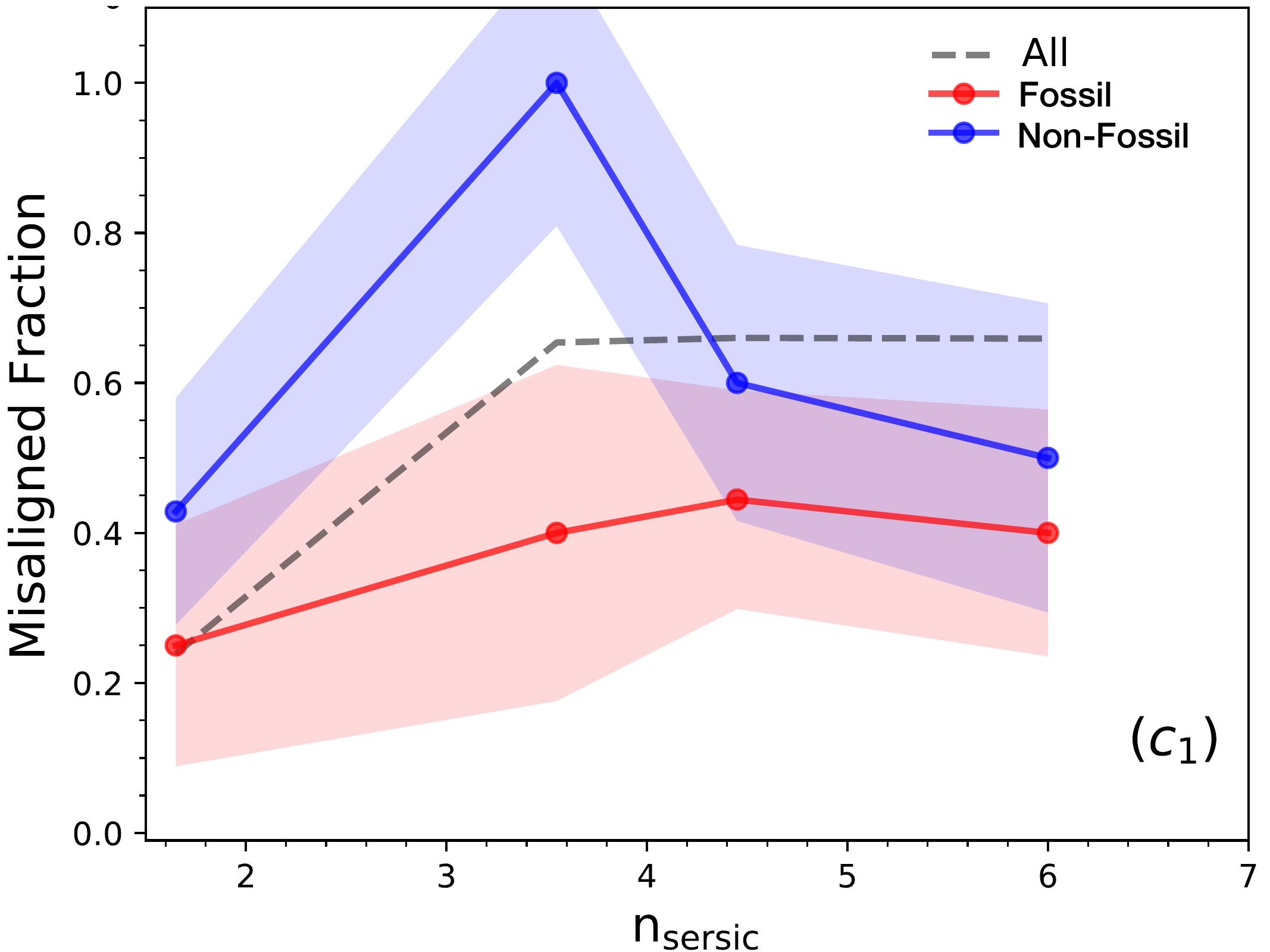}
    \caption{\textbf{Left:} The fraction of star - gas misalignment angel ($\delta PA$ > 30 degree) as a
function of stellar mass for BGGs hosted by relxed (red) and non-fossil (blue) groups. Dashed lines show the median trend for all BGGs in our sample ($b_{1}$). The inset figure ($b_{12}$) shows the median fraction of regular rotation BGGs as a function of stellar mass using the moving average method where the axes range is the same as in the main plot.
\textbf{Right:}  The fraction of star - gas misalignment angel as a function of BGGs Sérsic index ($c_{1}$), $n_{Sersic}$, with the same description as given in the left panel. In each panel, the errors (color-shaded lines) are $1\sigma$ confidence intervals on the fractions calculated using the bootstrap method. }
    \label{fig:5}
\end{figure}

Figure \ref{fig:5} ($b_1$ and $c_1$) shows  the fraction of misaligned BGGs at a given stellar mass and Sérsic index for sample of BGG in fossil versus non-fossil groups using Sydney-Australian-Astronomical-Observatory Multi-object Integral-Field Spectrograph (SAMI) galaxy survey \cite{Raouf2021}. In the left panel, for a given stellar mass, the fraction of misaligned \footnote{In our analysis, we took into account a one-sigma uncertainty in the catalog, which corresponds to a maximum error of 30 degrees for the PAs. The PAs were measured in an anticlockwise direction, with 0 degrees representing North (y = 0) in the sky. It is important to note that PAs with uncertainty values close to 30 degrees indicate less reliable fits. However, we found that such cases occurred in less than 5\% of all the SAMI samples, indicating that the majority of the fits were relatively robust. Therefore, we considered misalignment PAs greater than 30 degrees.} gas-star systems is primarily lower ($\sim$ 30\% less in all masses) in the fossil group's BGG compared to the BGG in the non-fossil sample and the median value across all galaxies. In the right, the non-fossil groups tend to have a higher fraction of misaligned BGGs at fixed Sersic index, again verifying the role of group dynamical state on the BGG dynamics. These differences are likely attributed to the extended period of tranquility experienced by BGGs in fossil groups, allowing them to settle and relax following earlier mergers and interactions.
BGGs in fossil groups, which are characterized by a more stable and settled environment, exhibit  $\sim$35\% more regular rotation fields and around 10\% more higher fractions of slow rotators compare to BGG in unrelaxded groups. 
Study of \cite{vandevoort2015} employed cosmological zoom-in simulations to carefully investigate the influence of mergers on the misalignment of gas discs in early-type galaxies. The research demonstrated that a significant merger could give rise to the formation of a new gas disc that is misaligned with the stellar rotation of the galaxy and persists for several gigayears.  

\section{Gas Misalignment in AGN-Dominated Galaxies: Insights from Hydrodynamic Simulations}

Molecular observations of the gas enveloping AGNs have uncovered a intricate scenario in which the observational signatures of mechanical and radiative feedback are confounded by the probable existence of starburst activity \cite{Aalto2012,Martin2011,Aladro2013,Watanabe2014}. Galaxies exhibiting both an AGN and robust circumnuclear star formation are particularly intriguing, as they encompass a intricate amalgamation of energetic processes, such as gas accretion and external infall, which contribute to a complex geometry and kinematics \cite{Knapen2019,Winkel2022,Imanishi2018}. These combined factors give rise to a multifaceted and dynamic system.
The ISM gas within the circumnuclear disk (CND) of an AGN-dominated galaxy, particularly a Seyfert-2 type galaxy like NGC 1068 \footnote{NGC 1068, discussed in this paper, is a spiral galaxy with a Seyfert nucleus. It is important to acknowledge that NGC 1068 is most representative of AGN activity in BGGs of relatively low-mass groups. The well-studied X-ray luminous groups, for which our knowledge is most complete, typically exhibit early-type BGGs with FR-I radio galaxies that are considerably more radio luminous than Seyferts. We also acknowledge the multifaceted nature of AGN activity and its various manifestations. However, we focus on the investigation of mechanical AGN feedback and its impact on the dynamics of gas within an idealized disc surrounding a supermassive black hole, using hydrodynamic simulations.}, was extensively investigated through hydrodynamic simulations referred to as HDGAS\footnote{https://github.com/mojtabaraouf/HDGAS} \cite{Raouf2023}. The study utilized these simulations to explore the properties and dynamics of the gas in the CND influenced by mechanical AGN feedback.
The study incorporated mechanical feedback from the AGN and employed the CHIMES non-equilibrium chemistry network to compare models with and without AGN feedback. The results revealed that the presence of an AGN significantly enhances CO formation in dense, clumpy regions surrounding supermassive black holes. This phenomenon is further elucidated by the examination of the counter-rotating gas disk in NGC 1068, as evidenced by kinematic maps of CO line emission. Further, in the models incorporating AGN feedback, specific regions of CND show higher velocity dispersion, represented by the parameter $\sigma$, which can be attributed to the superposition of velocity components associated with the outflow phenomenon.

\begin{figure}
    \centering
    \includegraphics[width=0.6\linewidth]{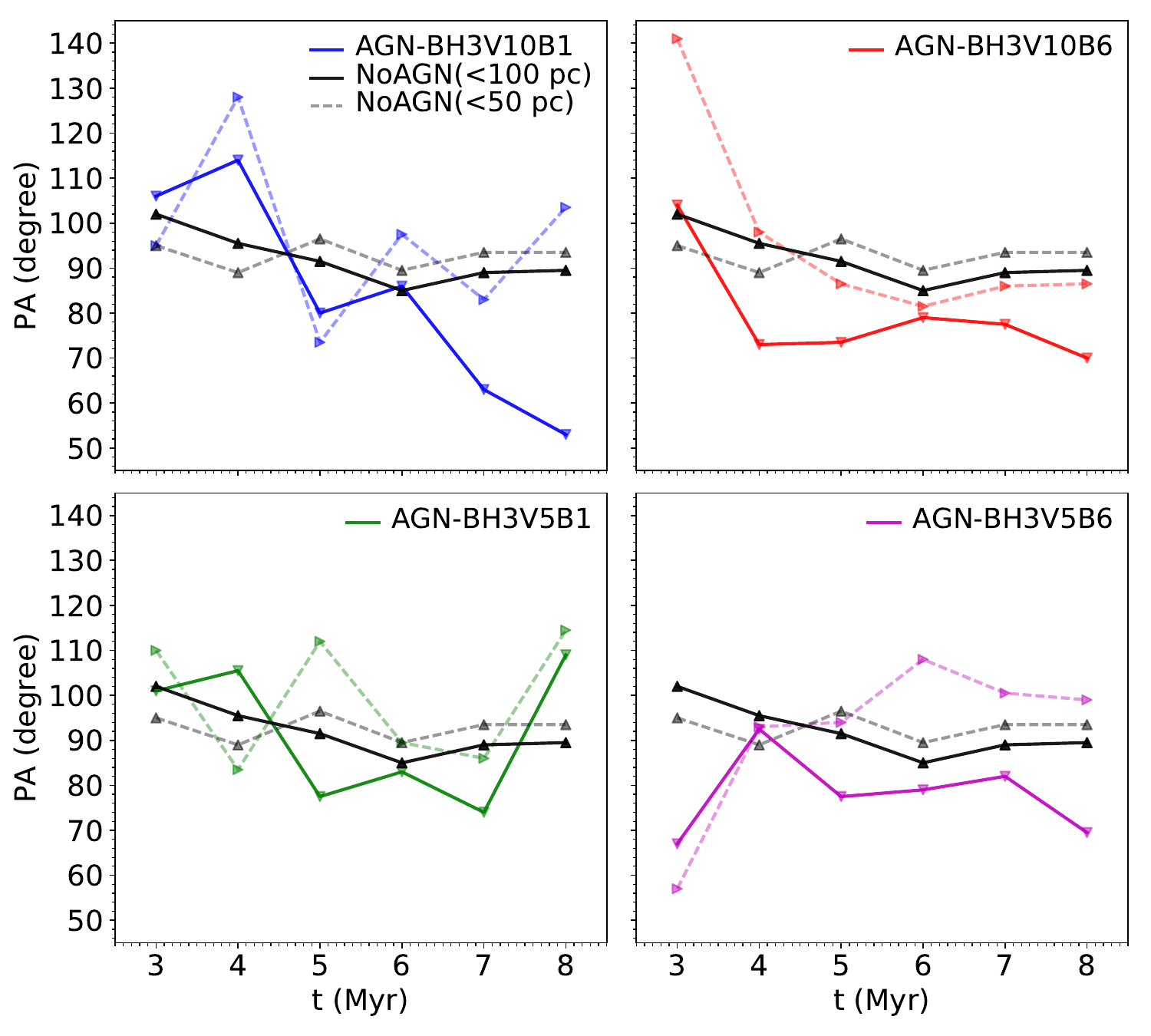}
    \caption{Kinematic position angles (PA) of CO(1-0) moment-1 maps within the central 50 (dashed lines) and 100 (solid lines) pc of the disk for AGN models with different energy and momentum loading factors and the model NoAGN (black -lines) measured at times between 3-8 Myr. A large proportion of such PA offsets occur within the
central r=50 pc of the disk surrounding the SMBH in low massloading
wind models (BH3V10B1, BH3V5B1), where the change in PA occurs within a shorter timeframe. There is a significant early offset (at 3-4 Myr) in models with a high mass-loading factor (BH3V10B6, BH3V5B6), followed by the steady PA trend to the
later times. Moreover, different scale radii of AGN models with low
velocity winds (BH3V5B6) exhibit a higher PA offset. The different
radii reveal similar trends for all models except for the model with
high wind velocity (i.e., BH3V10B1), which exhibits a large PA
offset at a later time (t$\sim$8 Myr) compared to an earlier times.}
    \label{fig:6}
\end{figure}

Figure \ref{fig:6} illustrates the variation in the disk's PA at various CND radii (<50 pc and <100 pc around SMBH) for four distinct AGN feedback models (high/low wind velocity and mass loading, see table 1 in \cite{Raouf2023}), each differing in energy and momentum loading compared to the model without AGN feedback.
The momentum and energy loading driven by the wind's velocity play a crucial role in maintaining the disk's misalignment at a scale of 100 pc over extended periods. 
The wind-induced effects lead to a persistent misalignment of the disk, characterized by a high offset in PA on the 100 pc scale, demonstrating a continued state of misalignment (top-left, AGN-BH3V10B1). However, in the bottom panels, when the wind velocity is lower (AGN-BH3V5B1, AGN-BH3V5B6), the disk appears to realign within a relatively short period of time. Increasing the mass-loading results in the smoothing out of the disk, following an initial substantial offset (top-right panel, AGN-BH3V10B6). Notably, in the top right panel, during earlier timeframes (around 3 to 4 million years), when both mass loading and wind velocity are high, the dynamics are predominantly influenced by the inflow phase. In contrast, the NoAGN model does not exhibit a significant PA offset at any time or radius.
We note that our study, conducted through hydrodynamic simulations, was not limited to specific environments and encompassed AGN-dominated galaxies across diverse conditions. The main focus of our investigation was to explore the effects of mechanical AGN feedback. Specifically, we examined the impacts of different jet velocities and momentum loading outflows on the dynamics of gas within an idealized disc surrounding a supermassive black hole. To gain insights into the role of AGN feedback, we compared the results with a model that did not incorporate any AGN feedback mechanisms.

\section{Discussion}

In this review, we have gained valuable insights into the impacts of black hole activity on the kinematics of gas in galaxies with different merger histories. By utilizing state-of-the-art cosmological simulations and observational surveys, we have explored various dimensions of this complex interplay. We start by examining the dynamics of galaxy groups and their formations states. It has been found that the BGGs in fossil systems, which are characterized by stable and settled environments, exhibit highly efficient black hole growth. The consumption of gas during the early stages of halo formation significantly contributes to the growth of black holes in these fossil systems (Figure \ref{fig:BH accretion rate&mass}).
Furthermore, we have investigated the growth history of both fossil and non-fossil halos and observed a significant difference in the timing of their peak merging activities (Figure \ref{fig:2}). In dynamically fossil environments, the BGGs exhibit lower levels of radio activity (see Figure \ref{fig:3}), indicating a relatively quiet phase of AGN activity. On the other hand, BGGs in non-fossil groups, experiencing more dynamic and turbulent environments, tend to have higher levels of star formation rates (Figure \ref{fig:4}). The occurrence of mergers in non-fossil groups contributes to the observed differences in the stellar populations and colors of BGGs.
A notable finding from our study is that the gas-star misalignment fraction in the BGGs of fossil groups is predominantly lower compared to both the BGGs in the non-fossil sample and the median value of all galaxies (Figure \ref{fig:5}). This suggests that the alignment between gas and stars in the BGGs is more preserved in fossil environments. 

By examining idealized gas discs surrounding SMBHs, our study demonstrates that the influence of winds on galaxies results in a persistent misalignment of the galactic disk.
The interaction between the galactic disk and the wind can disrupt the alignment of gas and stars, leading to changes in the kinematical major axis PA (Figure \ref{fig:6}). It is important to note that the spatial orientation of a galactic disk is typically described by two angles: the PA and the inclination angle. The PA represents the angle of the major axis of the disk on the sky, while the inclination angle describes the tilt of the disk with respect to the line of sight. In the case of net circular gas motion, the PA can provide information about the line of nodes, which is the intersection between the galactic disk and the plane of the sky. However, in galaxies influenced by complex gas dynamics, such as those affected by winds, the PA may not solely reflect the line of nodes. Factors like radial gas motions induced by winds can cause deviations in the PA from the disk's overall orientation. These radial gas motions, although induced by winds, can still have an impact on the angular momentum distribution and result in shifts in the disk's orientation. Therefore, the PA turn induced by winds can be considered as a significant influence on the orientation of the galactic disk.

In line with recent studies, we have found that BGGs in fossil groups with lower black hole activity display less kinematic misalignment. This finding supports the notion that external accretion events, such as galaxy mergers or the accretion of gas from the immediate environment, play a significant role in disrupting gas kinematics \cite{raimundo2023}. Misalignment between ionized gas and stellar kinematic angles is more prevalent in galaxies with a higher fraction of active black holes. Studies have also demonstrated the stable misalignment of gas discs after mergers \cite{vandevoort2015} and the influence of mechanical AGN feedback \cite{Raouf2023} on the gas misalignment.
To further advance our understanding, future research should focus on exploring the sources and dynamics of gas accretion in BGGs, particularly the roles of hot and cold accretion processes and the influence of active galactic nuclei (AGN). Larger samples of BGGs and the integration of data on stellar kinematics will strengthen our understanding of misalignment patterns. Additionally, investigating the temporal dynamics of gas-star realignment within galaxies is crucial, as it may mirror the evolutionary timescales of galaxy groups. This synchronized pattern of internal and external galactic evolution warrants further exploration.

By shedding light on the factors influencing BGG kinematics, this review lays the groundwork for future inquiries into the intricate processes governing the life cycles of galaxies and their group environments. The insights gained from this research will contribute to a more comprehensive understanding of the interplay between black hole activity, kinematic misalignment, and the environments of galaxies.

\vspace{6pt}

\dataavailability{Data from observation are publicly available in GAMA \href{http://www.gama-survey.org/}{(http://www.gama-survey.org/)} and SAMI \href{https://sami-survey.org/}{(https://sami-survey.org/)} galaxy surveyd. Data from the hydrodynamics simulations are based on the output from publicly available: (a) HDGAS simulation \citep{Raouf2023} following the GIZMO \cite{Hopkins2015} code at the following repository: \href{https://bitbucket.org/phopkins/gizmo-public/src/master/}{https://bitbucket.org/phopkins/gizmo-public/src/master/} and (b) Illustris-TNG simulation \href{www.tng-project.org}{(www.tng-project.org)}. Data from the semi-analytic model are based on the output from publicly availble model of  \href{https://github.com/mojtabaraouf/sage}{Radio-SAGE}. \textit{The data underlying this article will be shared on reasonable request to the corresponding author}.}



\acknowledgments{An Advanced Research Grant from the European Union (833460) supported this study as part of the  MOlecules as Probes of the Physics of EXternal (Moppex) galaxies project. FF would like to express her heartfelt appreciation to EuroSpaceHub and LUNEX EuroMoonMars Earth Space Innovation for their generous funding and unwavering support.}




\begin{adjustwidth}{-\extralength}{0cm}

\reftitle{References}


\bibliography{bibliography.bib}


\PublishersNote{}
\end{adjustwidth}
\end{document}